# Thermally-Driven Charge-Density-Wave Transitions in 1T-TaS$_2$ Thin-Film Devices: Prospects for GHz Switching Speed

Amirmahdi Mohammadzadeh[1], Saba Baraghani[1], Shenchu Yin[2], Fariborz Kargar[1], Jonathan P. Bird[2], and Alexander A. Balandin[1,*]

[1]Nano-Device Laboratory (NDL) and Phonon Optimized Engineered Materials (POEM) Center, Department of Electrical and Computer Engineering, University of California, Riverside, California 92521 USA

[2]Department of Electrical Engineering, University at Buffalo, the State University of New York, Buffalo, New York 14260 USA

**Abstract**

We report on the room-temperature switching of 1T-TaS$_2$ thin-film charge-density-wave devices, using nanosecond-duration electrical pulsing to construct their time-resolved current-voltage characteristics. The switching action is based upon the nearly-commensurate to incommensurate charge-density-wave phase transition in this material, which has a characteristic temperature of 350 K at thermal equilibrium. For sufficiently short pulses, with rise times in the nanosecond range, self-heating of the devices is suppressed, and their current-voltage characteristics are weakly non-linear and free of hysteresis. This changes as the pulse duration is increased to ~200 ns, where the current develops pronounced hysteresis that evolves non-monotonically with the pulse duration. By combining the results of our experiments with a numerical analysis of transient heat diffusion in these devices, we clearly reveal the thermal origins of their switching. In spite of this thermal character, our modeling suggests that suitable reduction of the size of these devices should allow their operation at GHz frequencies.

**Keywords:** charge-density-wave devices; van der Waals materials; 2D materials; pulsed electrical switching, 1T-TaS$_2$; transient Joule heating

[*] Corresponding author (A.A.B.): balandin@ece.ucr.edu ; web-site: http://balandingroup.ucr.edu/

Charge-density wave (CDW) phenomena have recently witnessed renewed interest, particularly in the context of two-dimensional (2D) van der Waals materials.[1–8] This interest is driven by both exciting physics, and possible practical applications, that can be derived from such materials. CDW transitions can be induced by various perturbations, including heating,[9] doping,[10–14] electrical pulsing,[15–17] substrate effects,[18] gate biasing,[17,19,20] and alteration of the thickness of the material.[21–23] The 1T polymorph of TaS$_2$ is one of the 2D van der Waals materials of the transition-metal dichalcogenide (TMD) group that reveals phase transitions in the form of abrupt resistivity changes and hysteresis.[24,25] Since two of these transitions are above room temperature (RT), 1T-TaS$_2$ has attracted attention for applications, such as high-speed memory devices,[17] oscillators,[26–28] transistor-less logic circuits,[29] oscillatory neural networks,[30] and logic gates.[31–33] The fact that two-terminal devices implemented with 1T-TaS$_2$ are radiation hard is an additional benefit.[27,34]

The resistivity changes exhibited by several TMDs, *e.g.* 1T-TaS$_2$ and 1T-TaSe$_2$, are attributed to transitions among different CDW phases.[24] The CDW phase is defined as a periodic modulation of the electronic charge density, accompanied by distortions in the underlying crystal lattice, and may be commensurate or incommensurate with the underlying crystal lattice.[16,35–43] The commensurate CDW (C-CDW) phase is typically observed at low temperature. It exhibits a stronger charge density distortion, and it is often pinned to the lattice.[44,45] The C-CDW phase only appears if the thickness of the films is sufficiently large. The lattice distortion in the C-CDW phase forms aggregates of 13 Ta atoms in the shape of a star in the basal plane, with 12 atoms at the vertices of the star inclines and 1 atom at its center.[46–50] From 180 K to 350 K, the aggregates partially melt, forming separate star-shaped islands that represent the nearly commensurate CDW (NC-CDW) phase. Increasing the temperature beyond 350 K, all the aggregate islands melt, and an incommensurate CDW (IC-CDW) phase emerges. Finally, the transition to the normal metallic phase (NP) happens in the range of 500-600 K.[17,20,58–61,24,51–57] The transition temperature between the NC and IC-CDW phases – at 350 K, conveniently above RT – makes 1T-TaS$_2$ particularly promising for device applications. The transition temperature is normally determined from transport measurements, in which the temperature of the sample is controlled externally while the electric bias is kept low to avoid local Joule heating.[62] While switching with the bias can pave the way to new electronic devices, different from conventional transistors, there are unanswered questions regarding the mechanism of the phase transitions in 2D CDW materials and devices.

In principle, at least, the application of an electric field can trigger the CDW phase transition directly, by affecting the charged ions and electrons in the crystal. If CDW-device operation relies on depinning of the CDW from defects, the electric field can set the CDW in motion above a certain threshold field. [63,64] Alternatively, however, the electric current induced by the field is responsible for Joule heating, which, in turn, can also trigger CDW phase transitions. The strength of the Joule heating should depend on details of the specific material, the device structure, and bias mode (DC vs. pulse) and duration. For these reasons, the mechanism of the different phase transitions that arise in different devices is still the subject of debate. [59,65–67] We have previously argued that, in 1T-TaS$_2$ devices on Si/SiO$_2$ substrates, the switching among different CDW phases that is caused by the application of an electric bias results from local Joule heating. [62] This conclusion was reached for devices for which the switching speeds were relatively slow, *i.e.* in the range of milliseconds to seconds. A recent study, on the other hand, which measured 1T-TaS$_2$ devices with electric pulses of duration 0.1 s, observed that Joule heating could be mostly inhibited. [53] Other related work, also based upon a pulsing technique, came to a different conclusion, namely that, even for pulses as short as ~0.1 ms, local Joule heating was not eliminated. [68] Elsewhere, a third possible outcome has also been reported, with the NC-CDW to IC-CDW transition being attributed to the electric-field action while the IC-CDW to NC-CDW transition was attributed to local Joule heating. [61] An extra factor that complicates the separation of field- and heating-driven effects arises from the difficulty of making a direct and accurate measurement of local temperature in the μm-scale channels of CDW devices.

Here, we report the results of an experimental investigation of 1T-TaS$_2$ devices under ultra-short electric pulses, combined with a computational study of heat dissipation in these structures. The 1T-TaS$_2$ CDW devices were fabricated on a highly resistive Si/SiO$_2$ substrate in order to allow for ns-duration electrical pulsing. We find that hysteresis, characteristic of the NC- to IC-CDW phase transition, emerges in the current-voltage characteristics of these devices at pulse durations in excess of ~200 ns, and that it exhibits a non-monotonic dependence with increasing pulse length. The experimentally observed evolution of the hysteresis, and our numerical solutions of the heat-diffusion equation, indicate that the switching is thermally driven. Despite this thermal nature, however, our modeling also suggests that downscaling of the device dimensions, and fine-tuning of the thermal resistance of the structure, should allow for fast operation of such devices, in the GHz frequency range.

High-quality single-crystal 1T-TaS$_2$ (HQ Graphene Co.) was used as the source material in our studies. Thin films of 1T-TaS$_2$ were mechanically exfoliated and transferred to a highly resistive (1-10 Ω-cm) Si/SiO$_2$ substrate (525-/0.3- µm, p-type, <100>) with the help of a transfer system that was built in-house. Uniform rectangular flakes were selected for device fabrication to match the pattern of the required on-chip co-planar waveguide (see figure 1e). The signal lines of this waveguide were 9-µm wide and were separated on either side by 3-µm gaps from two large ground planes, ensuring the 50-Ω matching needed for high-speed measurements.[69] To help preserve this matching, the 1T-TaS$_2$ crystals were chosen to be of the same width as the signal lines. Structures were fabricated by electron-beam lithography and lift-off of Ti/Au (20-/180-nm), deposited by electron-beam evaporation. To avoid degradation of the quasi-2D 1T-TaS$_2$ from exposure to chemicals and air, successive fabrication steps were performed with minimal delay. Fabricated devices were then stored and tested under vacuum conditions.

Figure 1a shows the temperature-dependent resistivity of a device implemented with a thin channel (<10 nm) of 1T-TaS$_2$, showing the NC to IC-CDW phase transition at ~350 K. The current – voltage (I-V) characteristics measured for the same 1T-TaS$_2$ device are presented in Figure 1b. The straight line is an example of the load-resistance characteristic required to make the oscillator shown schematically in Figure 1c. The implementation of such voltage-controlled oscillators from 1T-TaS$_2$ CDW devices has been reported previously by some of us. [20,27] Other fabricated devices had larger thicknesses (~70 nm) to make them more robust for the testing and to ensure contact quality. A schematic of the pulsed-measurement setup is presented in Figure 1d. This includes a pulse generator that sources ramped (triangular) current pulses of duration as short as 8 ns, and a mixed-signal oscilloscope (Keysight DSOX6004A, 6 GHz bandwidth) that is used to measure the time-dependent variation of the generated current.

The pulsed measurements were performed in a custom-designed setup with full impedance matching (50 Ω). The device chip was mounted on an FR-4 laminate carrier, which allowed the signal line of the previously mentioned on-chip coplanar waveguide to be connected to the measurement instruments, through semirigid coaxial cables and high-bandwidth SMA connectors. A 50-Ω RuO$_2$ thick film resistor, soldered onto the chip carrier, provided a matched termination for the input signal. An Aim-TTi TGP3151 pulse generator was used to provide the pulses. In this study, repetitive (75 kHz) triangular pulses of various width (8 – 13,333 ns) were applied to the CDW device input. The resulting output pulses were then captured at the 50-Ω input channel of the oscilloscope and correlated to the input signal to generate a pulsed I-V characteristic. All measurements were performed with the CDW devices mounted inside a customized, light-tight vacuum chamber (~10$^{-6}$ mbar), held at RT.

[Figure 1: Device schematic; microscopy image; measurement setup]

Figure 2 (a-d) shows the I-V characteristics obtained for different pulse durations (see Supplementary Figure S1). In all panels, only the I-V results of the first generated pulse are presented. This is essential to eliminate the influence of heat accumulation, originating from the application of repetitive pulses, on the I-V characteristics of the device. The most important observation is that, when the pulse duration is too short (see Figure 2a with the 8-ns pulse), the I-V curves do not reveal any hysteresis. As the pulse duration is increased, however, the hysteresis, associated with the NC to IC-CDW phase transition, emerges (Figure 2b, bias voltage of ~1.65 V). Visual inspection of the form of the I-V curves in this figure shows that the area associated with the hysteresis increases rather rapidly at first (compare Figures 2a and 2b), but then starts to decrease (Figure 2c and d). This behavior is attributed to the difference in the time constants for the rate of heat accumulation as a result of Joule heating and the rate of the heat dissipation to the underlying substrate. Overall, these data are consistent with the thermally-driven transition scenario; while the maximum strength of the electric field is the same for all of the pulses, the rate of the generated heat in the device and the dissipated heat depend strongly on the pulse duration. More specifically, we suggest that the ultra-short pulses of a few-ns duration are not sufficiently long to increase the local temperature to the vicinity of ~350 K, and to thereby drive the NC to IC-CDW phase transition. As the pulse duration is increased, however, heat should accumulate increasingly in the channel, leading to a rise of the local temperature and the emergence of the NC to IC-CDW hysteresis. The further evolution can then be understood as arising from an interplay of two time-

constants – one associated with how fast heat accumulates in the channel, and the other with how fast heat escapes from it. These time constants will depend on the pulse duration, the thermal resistance of the device, and its size. When the pulse becomes too long (of μs duration, as in Figure 2d) the local temperature does not decrease sufficiently between pulses to achieve full-size hysteresis, resulting in the decrease of the hysteresis window and saturation. Figure S2 exhibits the derivative of the I-V curves, $dI/dV$, shown in Figure 2. The phase transitions occur at bias voltages where abrupt changes in $dI/dV$ are observed.

[Figure 2: Experimental I-V characteristics with different current pulse durations]

To verify the proposed description of the switching mechanism, we analyzed the time-dependent heat dissipation in the device by solving the transient heat-diffusion equation. Calculations were performed for the specific layered structure of the experimentally tested devices, using the finite element method (COMSOL Multiphysics). The instantaneous Joule power generated in the channel is defined as $P(t) = V(t)I(t)$, where $V(t)$ is the applied potential, and $I(t)$ is the current in the 1T-TaS₂ channel.[62] The layered structure (see Figure 1-d) considered in the calculations includes the 0.5 mm-thick Si wafer, the SiO₂ gate dielectric, and the 1T-TaS₂ channel and metal electrodes. The through-plane thermal resistance of the Si/SiO₂ substrate includes three terms: $R_{th} = R_c + R_{SiO_2} + R_{Si}$, where $R_c$ is the thermal contact resistance between the 1T-TaS₂ film and the SiO₂, and $R_{SiO_2}$ and $R_{Si}$ are the thermal resistances of the SiO₂ and Si layers, respectively. The thermal resistance of the SiO₂ layer is defined as $R_{SiO_2} = t_{SiO_2}/(k_{SiO_2} \times A)$, where $t_{SiO_2}$ is the thickness (300 nm) of the SiO₂, $k_{SiO_2}$ is its thermal conductivity, and $A$ is the cross-sectional area of the channel. The interfacial contact resistance is $R_c = [G_{interface} \times A]^{-1}$ in which $G_{interface}$ is the interfacial thermal conductance and varies in a rather limited range for common 2D materials and substrates. For example, $G_{interface}$ for graphene, WTe₂, and MoS₂ with Si/SiO₂ substrates is 50, 33, and 15 MW/m²K, respectively.[70–73] Note that, for long current pulses, where the system is close to the steady-state condition, the average temperature rise in the device structure with SiO₂ layer rises to about $\Delta T \approx P \times R_{SiO_2}$, without even taking thermal contact resistance into account.[62]. The temperature rise driven by the interfacial contact resistance can be as high as 100-200 K.
6262,70,71

Figure 3 (a-b) present the results of simulations of the thermal profiles in a TaS$_2$ layer at the midpoint of 8 ns and 13335 ns pulses, respectively. The dimensions of the channel are $W \times L \times t_f = 7 \text{ μm} \times 3 \text{ μm} \times 140 \text{ nm}$, where $W$, $L$, and $t_f$ are the width, length, and the thickness, respectively. The Joule heat generated by the pulse mostly diffuses through the underlaying SiO$_2$ layer owing to the substantial difference in the in-plane and through-plane thermal resistance. The in-plane thermal conductivity of 1T-TaS$_2$ is almost three folds higher than that of SiO$_2$.[74,75] However, owing to the small thickness of the 1T-TaS$_2$ channel, the total in-plane thermal resistance of the TaS$_2$ active layer, defined as $R_{th,in-plane} = L/(k_f \times t_f \times W)$, is approximately two orders of magnitude higher than the total through-plane thermal resistance, $R_{th,through}$, of the underlying SiO$_2$/Si layers. Consequently, almost all the generated heat is dissipated through the Si/SiO$_2$ layer as seen in Figure 3. One can notice that the SiO$_2$ layer acts as an efficient thermal barrier owing to its small thermal conductivity and the thermal boundary resistance at the TaS$_2$-SiO$_2$ interface. While the heating is insufficient for 8-ns pulse duration (Figure 3a) to drive the NC- to IC-CDW phase transition, by increasing the pulse duration to 13.3 μs the local temperature in the TaS$_2$ rises well above 350 K (Figure 3b).

[Figure 3: Simulated temperature distribution along the cross-section of the device channel]

To further support our interpretation of the experimental results, we have simulated the I-V curves of the device tested experimentally (see Figure 4 and Supplementary Figure S3). The pulse duration in these simulations was chosen to match the different values used in the experiments. The pulses were implemented in the model by applying them at various predefined ramp rates. The current at each time step was then calculated using the equations embedded in the COMSOL electromagnetic module. Additional simulation details are provided in the Supplemental Information. One can clearly see from Figure 4 that the calculated trend for the evolution of the hysteresis is highly reminiscent of the experimentally measured one (see Figure 2). This is due to the temperature rise in the 1T-TaS$_2$ which causes a drastic decrease in resistance (Figure 1a) as a result of phase transition and thus, an increase in the amount of current that passes through the channel.

[Figure 4: Simulated pulsed measurement results at various pulse durations]

For a quantitative comparison of the experimental and modeled trends, we define the width of the hysteresis ($V_{max} - V_{min}$) as the difference in the voltage values observed during the up and down ramps of the voltage at a constant current. This width is a meaningful parameter as it defines the operational range of voltages for CDW voltage-controlled oscillators and other types of switches. Measured and calculated values of this parameter, determined for a current of $I_{SD} = 8$ mA, are presented in Figure 5. It is evident that both the experiment and simulations exhibit a peak at some particular value of the pulse duration, which should be defined by the interplay of the time constants for the generation and escape of heat. The absolute values of the measured and simulated hysteresis windows are different due to the fact that not all experimental factors, *e.g.* the heat dissipated to the environment, are accounted for in the simulation. The fine details of the simulation curve, i.e. deviations from a smooth trend, are numerical due to the relatively coarse mesh used.

[Figure 5: Experimental and simulated hysteresis window width]

Our experimental results show that for the tested device the hysteresis window appears for a pulse duration of ~200 ns. This corresponds to a CDW-device switching speed of ~5 MHz. From the discussion above, it is clear that the speed can be increased if one optimizes the device structure and decreases its size, to drive the desired phase transition by allowing heating to develop more rapidly. To demonstrate this point, in Figure 6 we have used our experimentally validated physics-based model to simulate the I-V curves of a device with channel length, width and thickness of 1 µm, 2 µm, and 10 nm, respectively. From this figure, one can see that the hysteresis appears at substantially shorter pulses than in Figure 2 and Figure 4. For a pulse duration of 20 ns, the calculated channel temperature reaches ~360 K (see Supplemental Figure S4), which is already above the NC to IC-CDW transition temperature. The hysteresis in fact begins to develop for pulses as short as ~1 ns (see Figure 6d), which corresponds to a CDW device speed of ~1 GHz. The obtained data therefore suggest that CDW devices implemented with quasi-2D 1T-TaS$_2$ channels can operate at high frequency, even if the NC-CDW to IC-CDW transition is thermally driven. We note that our study proves that this switching is achieved via thermal mechanism for the specific device design, and characteristic thermal resistance of the structure. This does not preclude the possibility of purely electrically driven CDW transitions in other device designs and structures. If the pure-electric field regime is achieved, our estimates for the device operational speed should increase to even higher values.

[Figure 6: Operation speed projection.]

In conclusion, we have reported on the room-temperature switching of 1T-TaS$_2$ thin-film CDW devices using nanosecond-duration electrical pulses. The switching action utilized the NC- to IC-CDW phase transition, which has a characteristic temperature of 350 K. The results of our rapid pulsed measurements, and numerical transient analysis of heat diffusion in the device structure, indicate the thermal nature of the switching in this type of CDW device. The modeling results suggest that a proper tuning of the device size and thermal resistance can allow these devices to operate at GHz frequencies, even when the switching is thermally driven.

**Acknowledgements**

The work at UC Riverside was supported, in part, by the U.S. Department of Energy under the contract No. DE-SC0021020 "Physical Mechanisms and Electric-Bias Control of Phase Transitions in Quasi-2D Charge-Density-Wave Quantum Materials". The fabrication of the CDW devices was performed at the UCR Nanofabrication Facility. The high-speed pulsed measurements were performed in Buffalo, under support from the U.S. Department of Energy, Office of Basic Energy Sciences, Division of Materials Sciences and Engineering (DE-FG02-04ER46180).

**Contributions**

A.A.B. conceived the idea, coordinated the project, contributed to experimental data analysis, and led the manuscript preparation; A.M. fabricated the devices, conducted current-voltage measurements, contributed to data analysis; S.B. conducted simulations and contributed to data analysis; F.K. contributed to analysis of experimental and simulation data and performed analytical derivations; S.Y. conducted pulse measurements. J.P.B. contributed to experimental data analysis. All authors contributed to writing the manuscript.

**The Data Availability Statement**

The data that support the findings of this study are available from the corresponding author upon reasonable request.

The supplemental information is available at the Applied Physics Letters journal web-site for free of charge.

**References**

[1] Q.H. Wang, K. Kalantar-Zadeh, A. Kis, J.N. Coleman, and M.S. Strano, Nat. Nanotechnol. **7**, 699 (2012).

[2] D. Costanzo, S. Jo, H. Berger, and A.F. Morpurgo, Nat. Nanotechnol. **11**, 339 (2015).

[3] F. Zhang, H. Zhang, S. Krylyuk, C.A. Milligan, Y. Zhu, D.Y. Zemlyanov, L.A. Bendersky, B.P. Burton, A. V Davydov, and J. Appenzeller, Nat. Mater. **18**, 55 (2019).

[4] M. Wuttig and N. Yamada, Nat. Mater. **6**, 824 (2007).

[5] J. Feng, R.A. Susilo, B. Lin, W. Deng, Y. Wang, B. Li, K. Jiang, Z. Chen, X. Xing, Z. Shi, C. Wang, and B. Chen, Adv. Electron. Mater. **6**, 1901427 (2020).

[6] M. Mahajan and K. Majumdar, ACS Nano **14**, 6803 (2020).

[7] T. Patel, J. Okamoto, T. Dekker, B. Yang, J. Gao, X. Luo, W. Lu, Y. Sun, and A.W. Tsen, Nano Lett. **20**, 7200 (2020).

[8] J. Yang, Y.Q. Wang, R.R. Zhang, L. Ma, W. Liu, Z. Qu, L. Zhang, S.L. Zhang, W. Tong, L. Pi, W.K. Zhu, and C.J. Zhang, Appl. Phys. Lett. **115**, 33102 (2019).

[9] J.A. Wilson, F.J. Di Salvo, and S. Mahajan, Adv. Phys. **24**, 117 (1975).

[10] Y. Yu, F. Yang, X.F. Lu, Y.J. Yan, Y.-H. Cho, L. Ma, X. Niu, S. Kim, Y.-W. Son, D. Feng, S. Li, S.-W. Cheong, X.H. Chen, and Y. Zhang, Nat. Nanotechnol. **10**, 270 (2015).

[11] F. Zwick, H. Berger, I. Vobornik, G. Margaritondo, L. Forró, C. Beeli, M. Onellion, G. Panaccione, A. Taleb-Ibrahimi, and M. Grioni, Phys. Rev. Lett. **81**, 1058 (1998).

[12] D.F. Shao, R.C. Xiao, W.J. Lu, H.Y. Lv, J.Y. Li, X.B. Zhu, and Y.P. Sun, Phys. Rev. B **94**, 125126 (2016).

[13] J.T. Ye, S. Inoue, K. Kobayashi, Y. Kasahara, H.T. Yuan, H. Shimotani, and Y. Iwasa, Nat. Mater. **9**, 125 (2010).

[14] R. Misra, M. McCarthy, and A.F. Hebard, Appl. Phys. Lett. **90**, (2007).

[15] D. Cho, S. Cheon, K.-S. Kim, S.-H. Lee, Y.-H. Cho, S.-W. Cheong, and H.W. Yeom, Nat.

Commun. **7**, 10453 (2016).

[16] L. Ma, C. Ye, Y. Yu, X.F. Lu, X. Niu, S. Kim, D. Feng, D. Tománek, Y.W. Son, X.H. Chen, and Y. Zhang, Nat. Commun. **7**, 10956 (2016).

[17] I. Vaskivskyi, I.A. Mihailovic, S. Brazovskii, J. Gospodaric, T. Mertelj, D. Svetin, P. Sutar, and D. Mihailovic, Nat. Commun. **7**, 11442 (2016).

[18] R. Zhao, Y. Wang, D. Deng, X. Luo, W.J. Lu, Y.P. Sun, Z.K. Liu, L.Q. Chen, and J. Robinson, Nano Lett. **17**, 3471 (2017).

[19] Y. Li, K.-A.N. Duerloo, K. Wauson, and E.J. Reed, Nat. Commun. **7**, 10671 (2016).

[20] G. Liu, B. Debnath, T.R. Pope, T.T. Salguero, R.K. Lake, and A.A. Balandin, Nat. Nanotechnol. **11**, 845 (2016).

[21] R. Samnakay, D. Wickramaratne, T.R. Pope, R.K. Lake, T.T. Salguero, and A.A. Balandin, Nano Lett. **15**, 2965 (2015).

[22] M. Yoshida, Y. Zhang, J. Ye, R. Suzuki, Y. Imai, S. Kimura, A. Fujiwara, and Y. Iwasa, Sci. Rep. **4**, 7302 (2015).

[23] P. Goli, J. Khan, D. Wickramaratne, R.K. Lake, and A.A. Balandin, Nano Lett **12**, (2012).

[24] B. Sipos, A.F. Kusmartseva, A. Akrap, H. Berger, L. Forró, and E. Tutǐ, Nat. Mater. **7**, 960 (2008).

[25] P.C. Börner, M.K. Kinyanjui, T. Björkman, T. Lehnert, A. V Krasheninnikov, and U. Kaiser, Appl. Phys. Lett. **113**, 173103 (2018).

[26] G. Liu, B. Debnath, T.R. Pope, T.T. Salguero, R.K. Lake, and A.A. Balandin, Nat. Nanotechnol. **11**, 845 (2016).

[27] G. Liu, E.X. Zhang, C.D. Liang, M.A. Bloodgood, T.T. Salguero, D.M. Fleetwood, and A.A. Balandin, IEEE Electron Device Lett. **38**, 1724 (2017).

[28] A.K. Geremew, S. Rumyantsev, B. Debnath, R.K. Lake, and A.A. Balandin, Appl. Phys. Lett. **116**, 163101 (2020).

[29] A.G. Khitun, A.K. Geremew, and A.A. Balandin, IEEE ELECTRON DEVICE Lett. **39**, 1449

(2018).

[30] A. Khitun, G. Liu, and A.A. Balandin, IEEE Trans. Nanotechnol. **16**, 860 (2017).

[31] Y. Wu, M. Bao, A. Khitun, J.Y. Kim, A. Hong, and K.L. Wang, J. Nanoelectron. Optoelectron. **4**, 394 (2009).

[32] A. Khitun and K.L. Wang, Superlattices Microstruct. **38**, 184 (2005).

[33] A. Khitun, M. Bao, and K.L. Wang, IEEE Trans. Magn. **44**, 2141 (2008).

[34] A.K. Geremew, F. Kargar, E.X. Zhang, S.E. Zhao, E. Aytan, M.A. Bloodgood, T.T. Salguero, S. Rumyantsev, A. Fedoseyev, D.M. Fleetwood, and A.A. Balandin, Nanoscale **11**, 8380 (2019).

[35] B. Sipos, A.F. Kusmartseva, A. Akrap, H. Berger, L. Forró, and E. Tutiš, Nat. Mater. **7**, 960 (2008).

[36] M.J. Hollander, Y. Liu, W.-J. Lu, L.-J. Li, Y.-P. Sun, J.A. Robinson, and S. Datta, Nano Lett. **15**, 1861 (2015).

[37] Y. Yu, F. Yang, X.F. Lu, Y.J. Yan, Y.-H. Cho, L. Ma, X. Niu, S. Kim, Y.-W. Son, D. Feng, S. Li, S.-W. Cheong, X.H. Chen, and Y. Zhang, Nat. Nanotechnol. **10**, 270 (2015).

[38] L. Stojchevska, I. Vaskivskyi, T. Mertelj, P. Kusar, D. Svetin, S. Brazovskii, and D. Mihailovic, Science (80-. ). **344**, 177 LP (2014).

[39] B. Burk, R.E. Thomson, J. Clarke, and A. Zettl, Science (80-. ). **257**, 362 (1992).

[40] X.L. Wu and C.M. Lieber, Science (80-. ). **243**, 1703 (1989).

[41] J.M. Carpinelli, H.H. Weitering, E. Ward Plummer, and R. Stumpf, Nature **381**, 398 (1996).

[42] S. Vogelgesang, G. Storeck, J.G. Horstmann, T. Diekmann, M. Sivis, S. Schramm, K. Rossnagel, S. Schäfer, and C. Ropers, Nat. Phys. **14**, 184 (2018).

[43] G. Grüner, Rev. Mod. Phys. **60**, 1129 (1988).

[44] J.A. Wilson, F.J. Di Salvo, and S. Mahajan, Phys. Rev. Lett. **32**, 882 (1974).

[45] Y.D. Wang, W.L. Yao, Z.M. Xin, T.T. Han, Z.G. Wang, L. Chen, C. Cai, Y. Li, and Y. Zhang, Nat. Commun. **11**, 1 (2020).

[46] P. Fazekas and E. Tosatti, Phys. B+C **99**, 183 (1980).

[47] N.F. MOTT, Rev. Mod. Phys. **40**, 677 (1968).

[48] J.A. Wilson, Phys. Rev. B **17**, 3880 (1978).

[49] A. Spijkerman, J.L. de Boer, A. Meetsma, G.A. Wiegers, and S. van Smaalen, Phys. Rev. B **56**, 13757 (1997).

[50] P. Fazekas and E. Tosatti, Philos. Mag. B Phys. Condens. Matter; Stat. Mech. Electron. Opt. Magn. Prop. **39**, 229 (1979).

[51] I. Vaskivskyi, J. Gospodaric, S. Brazovskii, D. Svetin, P. Sutar, E. Goreshnik, I.A. Mihailovic, T. Mertelj, and D. Mihailovic, Sci. Adv. **1**, e1500168 (2015).

[52] K. Sun, S. Sun, C. Zhu, H. Tian, H. Yang, and J. Li, Sci. Adv. **4**, eaas9660 (2018).

[53] M. Yoshida, T. Gokuden, R. Suzuki, M. Nakano, and Y. Iwasa, RAPID Commun. Phys. Rev. B **95**, 121405 (2017).

[54] M.J. Hollander, Y. Liu, W.J. Lu, L.J. Li, Y.P. Sun, J.A. Robinson, and S. Datta, Nano Lett. **15**, 1861 (2015).

[55] L. Stojchevska, I. Vaskivskyi, T. Mertelj, P. Kusar, D. Svetin, S. Brazovskii, and D. Mihailovic, Science (80-. ). **344**, 177 (2014).

[56] M. Yoshida, R. Suzuki, Y. Zhang, M. Nakano, and Y. Iwasa, Sci. Adv. **1**, e1500606 (2015).

[57] A.W. Tsen, R. Hovden, D. Wang, Y.D. Kim, J. Okamoto, K.A. Spoth, Y. Liu, W. Lu, Y. Sun, J.C. Hone, L.F. Kourkoutis, P. Kim, and A.N. Pasupathy, Proc. Natl. Acad. Sci. **112**, 15054 (2015).

[58] S. Zheng, F. Liu, C. Zhu, Z. Liu, and H.J. Fan, Nanoscale **9**, 2436 (2017).

[59] R. Salgado, A. Mohammadzadeh, F. Kargar, A. Geremew, C.-Y. Huang, M.A. Bloodgood, S. Rumyantsev, T.T. Salguero, and A.A. Balandin, Appl. Phys. Express **12**, 37001 (2019).

[60] W. Wen, Y. Zhu, C. Dang, W. Chen, and L. Xie, Nano Lett. **19**, 1805 (2019).

[61] C. Zhu, Y. Chen, F. Liu, S. Zheng, X. Li, A. Chaturvedi, J. Zhou, Q. Fu, Y. He, Q. Zeng, H.J.

Fan, H. Zhang, W.-J. Liu, T. Yu, and Z. Liu, ACS Nano **12**, 11203 (2018).

[62] A.K. Geremew, S. Rumyantsev, F. Kargar, B. Debnath, A. Nosek, M.A. Bloodgood, M. Bockrath, T.T. Salguero, R.K. Lake, and A.A. Balandin, ACS Nano **13**, 7231 (2019).

[63] S. Brown and G. Grüner, Sci. Am. **270**, 50 (1994).

[64] R.E. Thorne, Phys. Today **49**, 42 (1996).

[65] T.E. Kidd, T. Miller, M.Y. Chou, and T.C. Chiang, Phys. Rev. Lett. **88**, 226402/1 (2002).

[66] M. Porer, U. Leierseder, J.M. Ménard, H. Dachraoui, L. Mouchliadis, I.E. Perakis, U. Heinzmann, J. Demsar, K. Rossnagel, and R. Huber, Nat. Mater. **13**, 857 (2014).

[67] H. Cercellier, C. Monney, F. Clerc, C. Battaglia, L. Despont, M.G. Garnier, H. Beck, P. Aebi, L. Patthey, H. Berger, and L. Forró, Phys. Rev. Lett. **99**, 146403 (2007).

[68] Y. Ma, Y. Hou, C. Lu, L. Li, and C. Petrovic, Phys. Rev. B **97**, 195117 (2018).

[69] H. Ramamoorthy, R. Somphonsane, J. Radice, G. He, C.-P. Kwan, and J.P. Bird, Nano Lett. **16**, 399 (2016).

[70] Z. Chen, W. Jang, W. Bao, C.N. Lau, and C. Dames, Appl. Phys. Lett. **95**, 161910 (2009).

[71] E. Yalon, Ö.B. Aslan, K.K.H. Smithe, C.J. McClellan, S. V. Suryavanshi, F. Xiong, A. Sood, C.M. Neumann, X. Xu, K.E. Goodson, T.F. Heinz, and E. Pop, ACS Appl. Mater. Interfaces **9**, 43013 (2017).

[72] K.F. Mak, C.H. Lui, and T.F. Heinz, Appl. Phys. Lett. **97**, 221904 (2010).

[73] M.J. Mleczko, R.L. Xu, K. Okabe, H.-H. Kuo, I.R. Fisher, H.-S.P. Wong, Y. Nishi, and E. Pop, ACS Nano **10**, 7507 (2016).

[74] M.D. Núñez-Regueiro, J.M. Lopez-Castillo, and C. Ayache, Phys. Rev. Lett. **55**, 1931 (1985).

[75] W. Zhu, G. Zheng, S. Cao, and H. He, Sci. Rep. **8**, 10537 (2018).

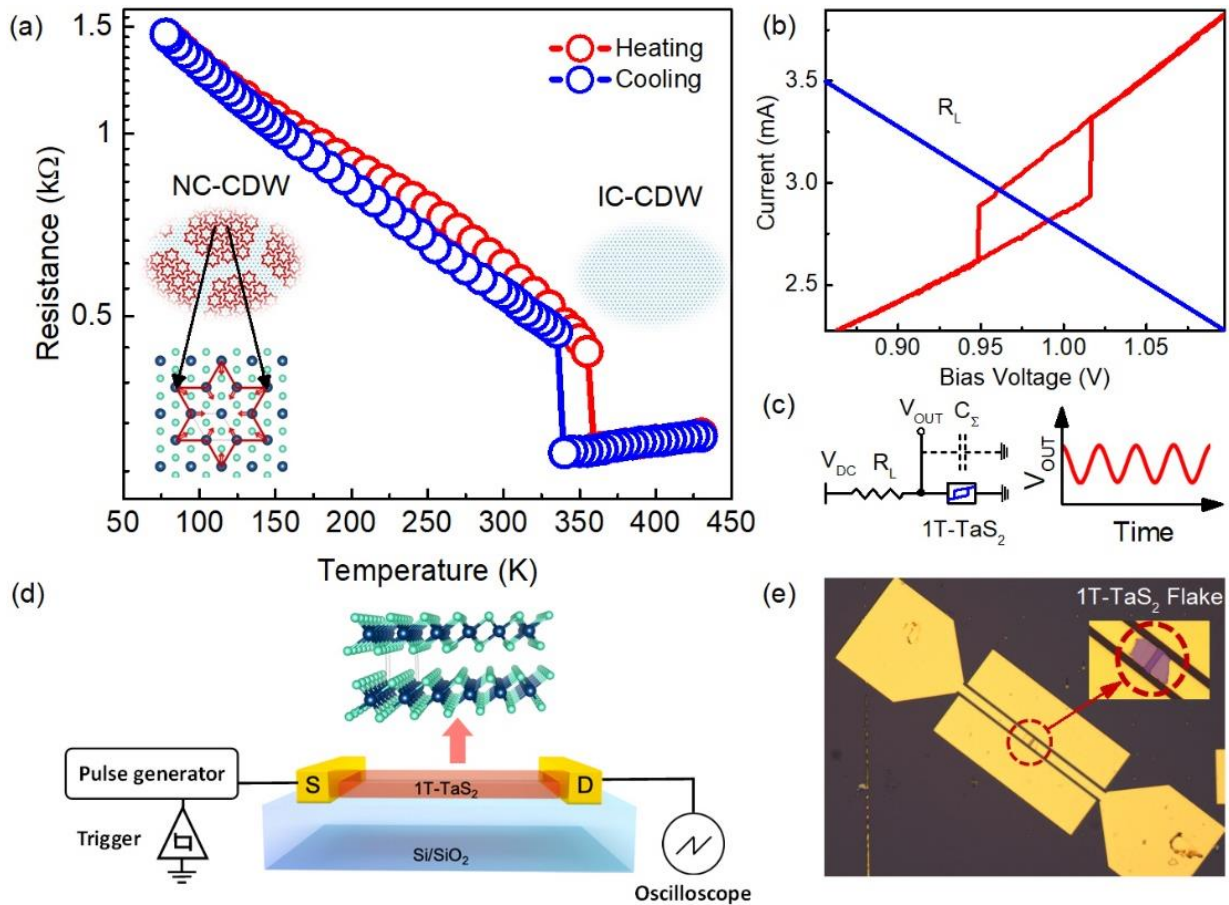

**Figure 1.** (a) Temperature-dependent resistance measurement for a thin (<10 nm) 1T-TaS$_2$ film showing the NC- to IC-CDW phase transition at ~350 K. As shown in the insets, the star-shaped islands in NC-CDW phase melt as the temperature rises above 350 K, resulting in the formation of the IC CDW and an accompanying decrease in resistance. (b) The typical current-voltage characteristic of a two-terminal 1T-TaS$_2$ device at RT, exhibiting the hysteresis window and the load line in CDW-based oscillator devices. (c) The left panel is the diagram of a 1T-TaS$_2$ oscillator circuit, including a 1T-TaS$_2$ thin film in series with a load resistor R$_S$ and a lumped capacitance C$_\Sigma$. The right panel shows a schematic of the real-time voltage oscillation generated by such devices. (d) Schematic showing the pulsed-measurement setup. A pulse generator creates repetitive current pulses as short as 8 ns. The generated current is then measured by a mixed-signal oscilloscope. (e) Optical image of the 1T-TaS$_2$ device designed for pulsed measurements. The inset shows a magnified image of the channel with the flake colored in purple for clarity. The gap between the two signal lines was 3 μm.

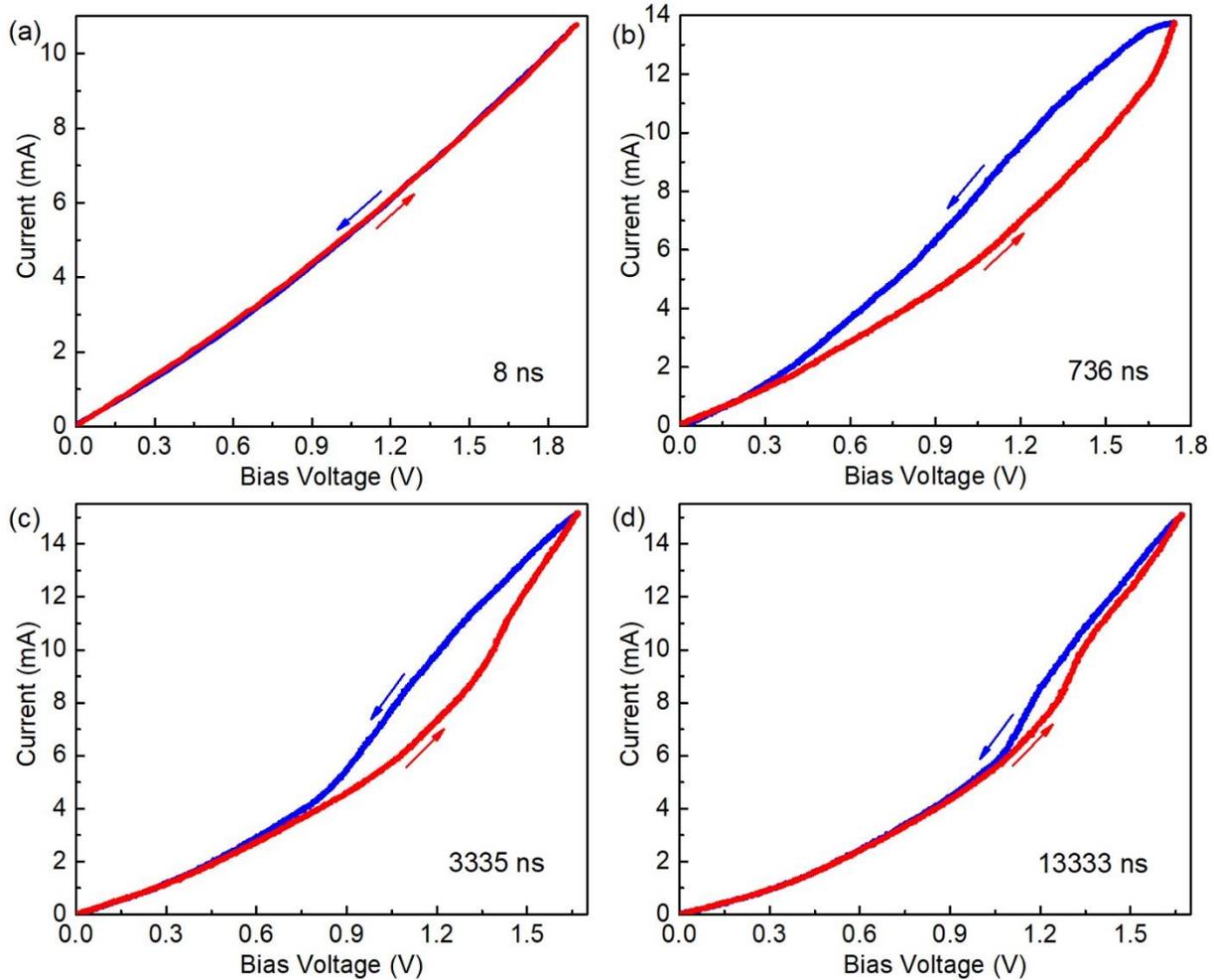

**Figure 2.** Experimental current-voltage characteristics, determined using total pulse durations of (a) 8 ns, (b) 736 ns, (c) 3,335 ns, and (d) 13,333 ns. For the shortest duration shown (8 ns), no hysteresis window is observed. With increasing the pulse duration, the width of the hysteresis window expands and then shrinks again as the pulse duration approaches to 13,333 ns. This behavior is attributed to the transient heat diffusion characteristics of the 1T-TaS$_2$ film, during the up and down sections of the pulse, causing the film to attain different temperatures at fixed bias in the hysteresis region.

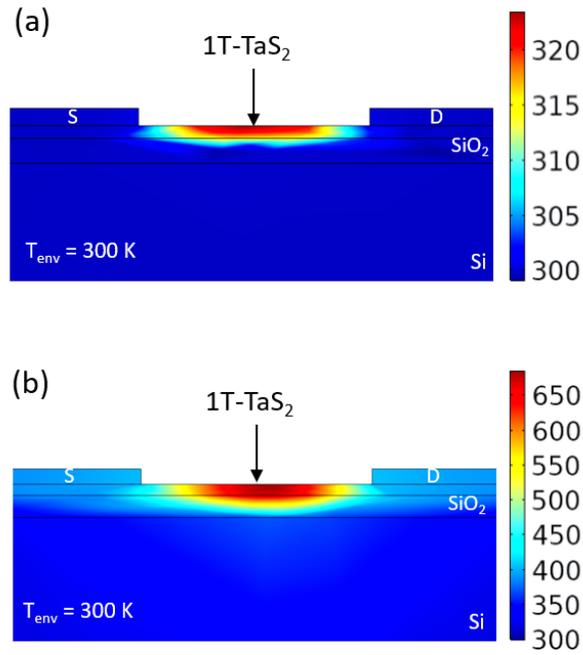

**Figure 3.** Cross-sectional view of the simulated temperature distribution near the device channel, for applied current pulses of duration (a) 8 ns, and (b) 13,335 ns. In case of the extremely short, 8-ns, pulses, the temperature of the film does not rise significantly above ambient, and remains below the NC- to IC-CDW phase transition temperature. For the much longer pulse, on the other hand, the temperature-rise is significant, exceeding this transition temperature.

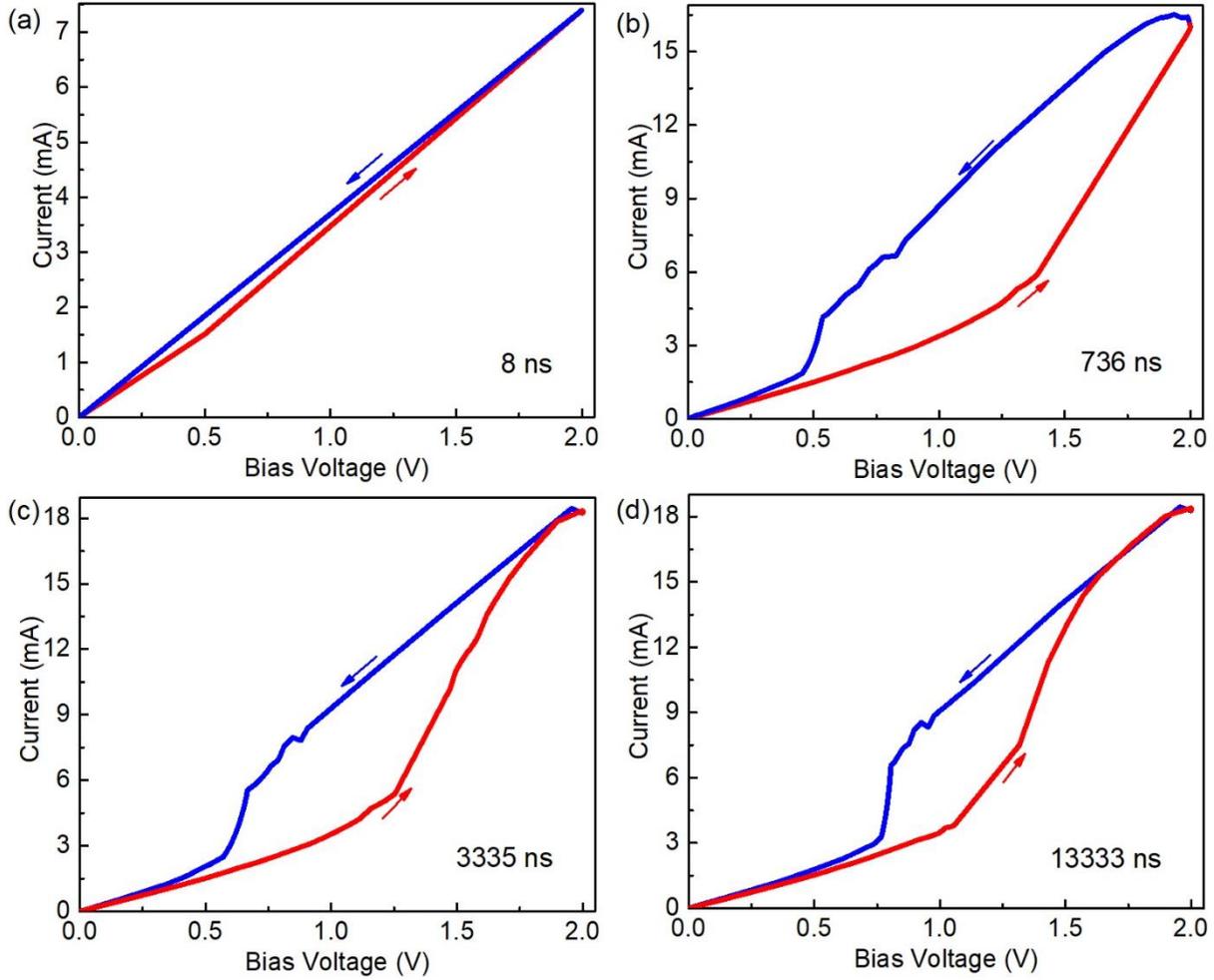

**Figure 4.** Simulated pulsed-measurement results at pulse durations of (a) 8 ns, (b) 736 ns, (c) 3,335 ns, and (d) 13,335 ns. The appearance, expansion and shrinkage of the hysteresis window follows the same trend as the experimental results. The ambient temperature is set at 300 K for these simulations.

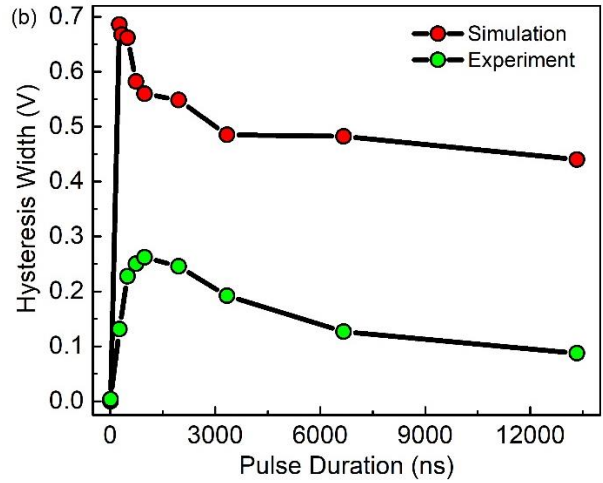

**Figure 5.** Experimental and simulated hysteresis window width ($V_{cooling}$-$V_{heating}$) calculated at the constant current of 8 mA and as a function of pulse duration. The experimental and theoretical results both follow the same trend, exhibiting a peak at shorter pulse durations and saturating at longer pulse times.

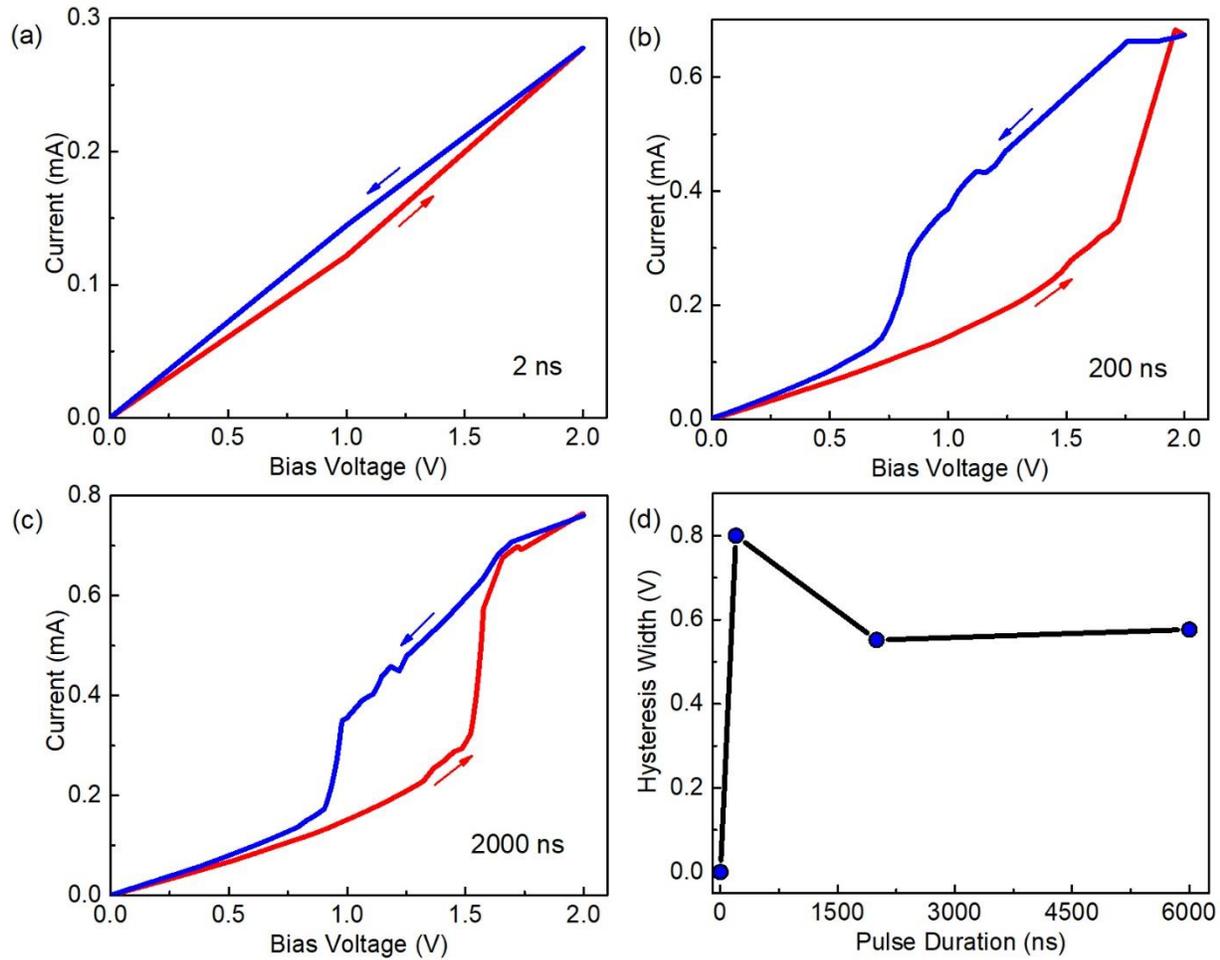

**Figure 6.** Simulated pulsed measurement results for the projected device with smaller dimensions (the device length, width and thickness are 1 µm, 2 µm, and 10 nm respectively), at various pulse durations: (a) 2 ns, (b) 200 ns, and (c) 2,000 ns. (d) Calculated hysteresis width for the same device as a function of pulse duration at constant current of 0.34 mA. The trend is similar to the previous experimental and simulation results. However, in this case, the hysteresis window can be observed even at smaller pulse durations. The hysteresis width has decreased compared to the larger devices. This confirms that, despite the dominant self-heating effect, careful tuning of the dimensions can lead to a device that can operate at GHz frequencies.